\documentclass[a4paper,english,british]{article}
\pdfoutput=1
\usepackage{lmodern}

\usepackage[T1]{fontenc}
\usepackage[utf8]{inputenc}
\usepackage{units}
\usepackage{mathrsfs}
\usepackage{amsmath}
\usepackage{amssymb}
\usepackage{graphicx}
\usepackage{esint}
\usepackage{overpic}
\usepackage{rotating}

\makeatletter

\pdfpageheight\paperheight
\pdfpagewidth\paperwidth

\usepackage{jcappub}

\makeatother

\usepackage{babel}
\begin{document}

\title{When matter matters}

\subheader{CERN-PH-TH/2013-074}

\author[a]{Damien A. Easson,}

\author[b]{Ignacy Sawicki}

\author[c,d]{and Alexander Vikman}

\affiliation[a]{Department of Physics \& School of Earth and Space Exploration \&
Beyond Center,\\
Arizona State University, Tempe, AZ, 85287-1504, USA\\
}

\affiliation[b]{Institut für Theoretische Physik, Ruprecht-Karls-Universität Heidelberg\\
Philosophenweg 16, 69120 Heidelberg, Germany\\
}

\affiliation[c]{CERN, Theory Division, CH-1211 Genève 23, Switzerland\\
}

\affiliation[d]{Stanford University, Department of Physics, Stanford, CA 94305,
USA\\
}

\emailAdd{easson@asu.edu}

\emailAdd{ignacy.sawicki@uni-heidelberg.de}

\emailAdd{alexander.vikman@cern.ch}

\abstract{We study a recently\emph{ }proposed\emph{ }scenario for the early
universe:\emph{ Sublumina}l\emph{ Galilean Genesis}. We prove that
without any other matter present in the spatially flat Friedmann universe,
the perturbations of the \emph{Galileon} scalar field propagate with
a speed at most equal to the speed of light. This proof applies to
\emph{all} cosmological solutions---to the \emph{whole phase space}.
However, in a more realistic situation, when one includes any matter
which is not directly coupled to the \emph{Galileon}, there always
exists a region of phase space where these perturbations propagate
\emph{superluminally, }indeed with \emph{arbitrarily high} speed. We
illustrate our analytic proof with numerical computations. We discuss
the implications of this result for the possible UV completion of
the model.}

\maketitle

\section{Introduction}

The theory of generation of quantum cosmological perturbations\footnote{For the scalar (or energy density) perturbations which have been already observed, see \cite{Mukhanov:1981xt,Chibisov:1982nx,Hawking:1982cz,Starobinsky:1982ee,Guth:1982ec,Bardeen:1983qw,Mukhanov:1985rz,Sasaki:1986hm,Mukhanov:2003xw,Mukhanov:2013tua}, while for the tensor modes (or gravitational waves), see an earlier work \cite{Starobinsky:1979ty}.}
during early-universe inflation\footnote{See  \cite{Guth:1980zm,Linde:1981mu,Albrecht:1982wi} and \cite{Starobinsky:1979ty,Starobinsky:1980te,Sato:1980yn,Kazanas:1980tx,Brout:1979bd,Gliner:1970} for earlier works on the quasi-de Sitter stage in the early universe. We are thankful to the anonymous referee for drawing our attention to works \cite{Gliner:1970,Sato:1980yn}. }
received additional strong support from the recent results of the
Planck collaboration \cite{Ade:2013lta}. Inflationary spacetimes
usually feature a singularity or in other words strong quantum gravity
stage far enough back in their past \cite{Borde:2001nh}. For a universe
with no spatial curvature, the only known method of avoiding a strong
quantum gravity stage is to posit a period in the evolution of the
universe where the Null Energy Condition (NEC)%
\footnote{This condition states that $T^{\mu\nu}n_{\mu}n_{\nu}\geq0$, where
$T^{\mu\nu}$ is an energy-momentum tensor $n^{\mu}$ represents all
null vectors. %
} is violated. However, until recently, such classical and local constructions
in the context of standard general relativity were always plagued
by some sort of pathological instabilities\textbf{---}either ghosts
or gradient instabilities or both, see e.g. \cite{Dubovsky:2005xd}.%
\footnote{One may try to avoid these problems by adding higher derivatives directly
to the EFT for \emph{perturbations} see e.g. \cite{Creminelli:2006xe,Creminelli:2007aq,Cheung:2007st,Creminelli:2008wc}.
This approach may be useful for a better systematisation of the perturbative
expansion but cannot elucidate the behaviour of the cosmological background. %
} The situation has changed with the rediscovery \cite{Luty:2003vm,Nicolis:2004qq,Nicolis:2008in,Deffayet:2009wt,Deffayet:2009mn,deRham:2010eu,Deffayet:2010qz,Deffayet:2011gz,Kobayashi:2011nu}
of more general single scalar-field theories \cite{Horndeski:1974},
which contain higher derivatives of the scalar field in the action
but still have only one scalar degree of freedom. These theories were
explicitly demonstrated to be able to violate the NEC with neither
ghost nor gradient instabilities in at least a part of their phase
space,%
\footnote{However, it does not mean that there are no instabilities at all.
Indeed, there are good physical arguments \cite{Sawicki:2012pz} that
there are more subtle and less dangerous run-away instabilities which
are still present in these theories. %
} see e.g.~\cite{Nicolis:2009qm,Creminelli:2010ba,Deffayet:2010qz,Kobayashi:2010cm}.
Moreover, it was shown in \cite{Easson:2011zy} that these theories
can realise a bounce and avoid a strong quantum-gravity regime but
only at the price of a strongly coupled regime of these non-renormalizable
theories in the initial, now collapsing, universe.

An interesting and unusual scenario for the initial stages of the
universe was proposed in \cite{Creminelli:2010ba}: \emph{Galilean
Genesis}. There, the universe starts from the Minkowski spacetime
which corresponds to an unstable and singular configuration of the
\emph{Galileon} scalar field with zero energy density. Small departures
from this original state violate the NEC. Thus starting from this
configuration the universe can expand by generating the energy density
during its evolution. In fact this evolution is just a superinflationary
stage with a very small acceleration.%
\footnote{This is also the case for other Genesis scenarios \cite{Hinterbichler:2012fr,Hinterbichler:2012yn}.%
} Nearly scale-invariant cosmological perturbations are supposed to
be generated by introducing an extra spectator scalar field coupled
to the \emph{Galileon }in such a way that effective metric reproduces
quasi de Sitter. Eventually the \emph{Galileon} should disappear by
transmitting its energy into radiation, before the universe runs into
a Big Rip singularity \cite{Starobinsky:1999yw,Caldwell:2003vq}.
This may happen through a reheating mechanism proposed in \cite{LevasseurPerreault:2011mw}.
However, it is not clear whether one can really get rid of the \emph{Galileon},
so that it does not spoil the late evolution of the universe, because
some of the \emph{Galileon's} configurations have negative energy
densities while on some other the NEC is broken so that the energy
density in them will still grow with time even after the end of reheating.
In \cite{Creminelli:2010ba}, only the particular \emph{Genesis }solution
was studied, while bouncing trajectories in the model were first investigated
in \cite{Qiu:2011cy}. The full cosmological phase space was investigated
in \cite{Easson:2011zy}, demonstrating that, without a reheating
mechanism, all trajectories eventually evolve to the configurations
where the sound speed becomes zero, so that the original effective
field theory (EFT) becomes infinitely strongly coupled. This strong-coupling
regime occurs before the would-be Big Rip future singularity. This
conclusion also applies to the \emph{Genesis} solution which turns
out to be a separatrix in the cosmological phase space of the \emph{Galilean
Genesis}. 

Perturbations propagating around the \emph{Genesis }solution initially
do so with a sound speed equal to that of light, slowing down as the
expansion becomes significant, and eventually vanishing. In these
higher-derivative scalar theories, superluminal propagation is common,
see e.g.~\cite{Adams:2006sv}. While its presence does not necessarily
signify the existence of causal paradoxes or problems on the level
of EFT see e.g.~\cite{Bruneton:2006gf,Kang:2007vs,Babichev:2007dw,Geroch:2010da,Burrage:2011cr}\footnote{For other recent studies of superluminality in various derivatively coupled theories see e.g. \cite{Evslin:2011vh,Evslin:2011rj,Hinterbichler:2009kq,Gruzinov:2011sq,deRham:2011pt,Deser:2012qx,deFromont:2013iwa}.},
it does mean that any EFT where a connected part of the phase space
features superluminality cannot have an ultra-violet (UV) completion
which is local, Lorentz invariant and corresponds to some weakly coupled
heavier particles \cite{Adams:2006sv}. Thus such theories cannot
have a standard Wilsonian UV completion, but may still be completed
in a non-standard way, via e.g.\emph{~classicalization} introduced
in \cite{Dvali:2010jz,Dvali:2010ns} for such derivatively coupled
scalars. Moreover, it was argued \cite{Dvali:2011nj,Dvali:2012zc,Vikman:2012bx}
that \emph{classicalization }may only work provided that superluminality
is possible. 

To avoid the superluminality of perturbations and give a hope for
a Wilsonian UV completion, in \cite{Creminelli:2012my},\emph{ Subluminal
Galilean Genesis }was introduced. An additional parameter in the action
ensures that the sound speed on the \emph{Genesis} trajectory can
be arbitrarily small, while the substantial features of the original
\emph{Galilean Genesis} are preserved. This should imply that in other
parts of the phase space, at least those close to the \emph{Genesis
}trajectory, the sound speed would be similarly reduced, in principle
removing any such potential superluminality.

In this paper, we extend our phase-space analysis presented in \cite{Easson:2011zy}
to the \emph{Subluminal Galilean Genesis }model. 

First, in section \ref{sec:RobSub} we analytically prove that any
configuration of a spatially-flat Friedmann universe filled solely
by the \emph{Galileon} with the action and parameters from \emph{Subluminal
Galilean Genesis }\cite{Creminelli:2012my} has a subluminal sound
speed. In fact, this is true even of the original \emph{Galilean Genesis}
model \cite{Creminelli:2010ba}. 

However, as a result of the imperfect nature of the fluid \cite{Pujolas:2011he}
described by the scalar field or kinetic mixing / braiding of the
scalar with the metric, the local properties of the \emph{Galileon}
do not just depend on the local scalar-field variables, but also on
the local Ricci tensor and, through the Einstein equations, on all
other external matter species present locally. In particular, this
implies that the sound speed of the \emph{Galileon} changes in the
presence of other matter species. Using this unusual property, we
show analytically in section \ref{sec:superlu} that for all admissible
parameter values of the \emph{Subluminal Galilean Genesis} model,
there are \emph{always} cosmological configurations which feature
superluminal propagation given the addition of an appropriate external
positive energy density with a normal equation of state $0<w<1$.
Moreover, perturbations around these configurations are not ghosts
and some of these configurations correspond to the same values of
the local scalar-field variables available in the phase space of the
\emph{Subluminal Galilean Genesis} without any external matter. In
particular, these configurations can be obtained by a continuous transformation
from the \emph{Genesis} trajectory. We illustrate our analytic proofs
with numerical computations. 

Thus as such, none of the \emph{Subluminal Galilean Genesis }models
is free of superluminality in the presence of external matter and
by the arguments of \cite{Adams:2006sv} cannot enjoy a standard Wilsonian
UV completion. 
\section{Model and Main Equations}

We study the class of models for the \emph{Galileon} scalar-field
$\pi$, minimally coupled to gravity $g_{\mu\nu}$, the dynamics of
which are given by the action introduced in \cite{Creminelli:2010ba,Creminelli:2012my}
\footnote{Contrary to \cite{Creminelli:2010ba,Creminelli:2012my} we use the
signature convention $\left(+,-,-,-\right)$, so that the sign in
front of the first term is ``wrong'' ---it corresponds to a ghost. %
}
\begin{equation}
S_{\pi}=\int\mbox{d}^{4}\! x\sqrt{-g}\left[-f^{2}e^{2\pi}\left(\partial\pi\right)^{2}+\gamma\,\frac{\beta}{2}\,\left(\partial\pi\right)^{4}+\gamma\left(\partial\pi\right)^{2}\Box\pi\right]\,,\label{eq:action}
\end{equation}
where $\Box=g^{\mu\nu}\nabla_{\mu}\nabla_{\nu},$ with $\nabla_{\mu}$
representing the covariant derivative while 
\[
\beta=1+\alpha\,,
\]
is a dimensionless parameter bound to be in the range $1\leq\beta<4$,
and another dimensionless parameter 
\begin{equation}
\gamma=\left(\frac{f}{\Lambda}\right)^{3}\gg1\,,\label{eq:gamma}
\end{equation}
is a combination of constants $f$ and $\Lambda$ of dimension one.
The positive parameter $\alpha$ was introduced in \cite{Creminelli:2012my}
and represents the only difference between the system (\ref{eq:action})
and the so-called \emph{Conformal Galileon} used to drive the \emph{Galilean
Genesis} scenario in \cite{Creminelli:2010ba}, where $\alpha=0$,
and which was also studied in context of a bouncing cosmology in e.g.
\cite{Qiu:2011cy,Easson:2011zy,Osipov:2013ssa}. The reason for the
introduction of the positive $\alpha$ in \cite{Creminelli:2012my}
is the ``robust subluminality'' around the \emph{Genesis} solution.
We will mostly use $\beta$ to simplify formulae. 

It is useful to perform the following field redefinition 
\begin{equation}
\pi=\ln\left(\frac{\phi}{f}\right)\,,\label{eq:new_field_phi}
\end{equation}
so that one obtains a form of the action where the scalar field $\phi$
has standard dimensions, 
\begin{equation}
S_{\phi}=\int\mbox{d}^{4}\! x\sqrt{-g}\left[-\left(\partial\phi\right)^{2}+\gamma\left(\frac{\beta-2}{2}\right)\frac{\left(\partial\phi\right)^{4}}{\phi^{4}}+\gamma\,\frac{\left(\partial\phi\right)^{2}}{\phi^{2}}\,\frac{\Box\phi}{\phi}\right]\,,\label{eq:action_phi}
\end{equation}
and the action has just two dimensionless parameters $\beta$ and
$\gamma$. In particular, from this form of the action, it is clear
that no single physical observable related purely to the evolution
of this system can depend on $\Lambda$ and $f$ separately but only
on their ratio given by $\gamma$. However, if there are \emph{non-minimal
}but \emph{non-universal} direct couplings to some external fields
/ matter through 
\begin{equation}
g_{\mu\nu}^{\text{matter}}=e^{2\pi}g_{\mu\nu}=\left(\frac{\phi}{f}\right)^{2}g_{\mu\nu}\,,\label{eq:g_matter}
\end{equation}
as was proposed in \cite{Creminelli:2010ba} for a spectator field
$\sigma$, then $f$ plays a role of the coupling constant to this
type of external matter, see \cite{Nicolis:2008in}. This coupling
is not only needed to generate fluctuations as in \cite{Creminelli:2010ba}
but also to reheat and possibly exit from the super-inflationary regime
\cite{LevasseurPerreault:2011mw}. Note that the coupling cannot be
universal for all matter, since the model (\ref{eq:action}) would
then just correspond to inflation written in a conformally rescaled
metric. Therefore there should be at least some matter species which
do not couple to the \emph{Galileon} directly through (\ref{eq:g_matter})
but only through gravity. Further we note that the system simplifies
for $\beta=2$. 

The dimensionless constant $\gamma$ can be moved in front of the
whole action if one rescales the coordinates as
\begin{eqnarray}
x^{\mu}\rightarrow\gamma^{1/2}\, y^{\mu}\,, & \mbox{so that} & S_{\phi}\rightarrow\gamma\, S_{\phi}\,.\label{eq:rescaling}
\end{eqnarray}
Note that the Einstein-Hilbert action scales in the same way $S_{\text{EH}}\rightarrow\gamma\, S_{\text{EH}}$.
Hence the dynamics of the \emph{Galileon} field $\phi$ interacting
with gravity is completely independent of $\gamma$ in the absence
of other fields. In a realistic universe there should be other matter
species, e.g. radiation or cold dark matter. If we assume that those
are not coupled to the \emph{Galileon }directly but only through gravity
one should rescale their energy density (and pressure) by 
\begin{equation}
\rho\rightarrow\rho/\gamma\,,\label{eq:Rho_rescaled}
\end{equation}
so that, after this rescaling, $\gamma$ disappears from all equations.
Therefore the rescaled energy density is allowed to be larger than
unity but should be parametrically smaller than $\gamma$ in order
to avoid physical transplanckian energy densities. 
\subsection{Some properties of models with Kinetic Gravity Braiding}

Both the versions of the action for \emph{Galilean Genesis, }(\ref{eq:action})
and (\ref{eq:action_phi}), belong to the same class %
\footnote{This class of models was introduced in \cite{Deffayet:2010qz} as
kinetic gravity braiding and then slightly later in \cite{Kobayashi:2010cm}. %
} 
\begin{equation}
S_{\varphi}=\int\mbox{d}^{4}x\sqrt{-g}\left[K\left(\varphi,X\right)+G\left(\varphi,X\right)\Box\varphi\right]\,,\label{eq:action_KGB}
\end{equation}
where $K$ and $G$ are arbitrary functions of $\varphi$ and where
\begin{equation}
X=\frac{1}{2}g^{\mu\nu}\,\partial_{\mu}\varphi\,\partial_{\nu}\varphi\,.\label{eq:X}
\end{equation}
We are going to work with both actions (\ref{eq:action}) and (\ref{eq:action_phi}),
hence, to make the paper self-contained, we will list the main dynamical
equations and stability criteria valid for a general theory of type
(\ref{eq:action_KGB}). After that one can use these equations by
substituting $\varphi\rightarrow\pi$ or $\phi$ and the corresponding
$K$ and $G$. All these equations can be found with derivations in
this form in \cite{Deffayet:2010qz,Pujolas:2011he}. In a cosmological
setup, for these systems, it is convenient to use hydrodynamical notation
and analogy introduced in \cite{Pujolas:2011he}. 

The \emph{diffusivity} measuring the kinetic mixing / braiding of
$\varphi$ with the metric is 
\begin{equation}
\kappa\equiv2XG_{,X}\,,\label{eq:diffusivity}
\end{equation}
where $\left(\;\right)_{,X}\equiv\partial\left(\;\right)/\partial X$.
The effective mass per shift-charge / chemical potential is 
\begin{equation}
m\equiv\partial_{t}\varphi\equiv\dot{\varphi}\,.\label{eq:chemical_Pot}
\end{equation}
The pressure of this scalar imperfect fluid is 
\begin{equation}
\mathcal{P}=K-m^{2}G_{,\varphi}-\kappa\dot{m}\,,\label{eq:pressure}
\end{equation}
while the density of shift charges (which are not conserved for the
systems studied in this paper) is 
\begin{equation}
n=K_{,m}-2mG_{,\varphi}+3H\kappa\,,\label{eq:number_density}
\end{equation}
where $H$ is the Hubble parameter. The corresponding energy density
is 
\begin{equation}
\mathcal{E}=mn-\mathcal{P}-\kappa\dot{m}=\varepsilon+3Hm\kappa\,,\label{eq:total_energy_density}
\end{equation}
where 
\begin{equation}
\varepsilon=m\left(K_{,m}-mG_{,\varphi}\right)-K\,,\label{eq:perfect_part_of_energy}
\end{equation}
is a ``perfect part'' of the energy density which is independent
of the expansion or Hubble parameter. In the spatially flat Friedmann
universe we have %
\footnote{Throughout the paper we use the reduced Planck units where $M_{\text{Pl}}=\left(8\pi G_{\text{N}}\right)^{-1/2}=1$.%
} 
\begin{equation}
H^{2}=\frac{1}{3}\left(\mathcal{E}+\rho\right)=\kappa m\, H+\frac{1}{3}\left(\varepsilon+\rho\right)\,,\label{eq:Friedmann_1}
\end{equation}
where $\rho$ is the energy density in external matter species. We
have assumed that these species are not coupled to the \emph{Galileon}
directly, so that the corresponding equation of motion is the continuity
equation $\dot{\rho}+3H\left(\rho+p\right)=0$, where $p$ is the
their pressure. The second Friedmann equation is 
\begin{equation}
\dot{H}=-\frac{1}{2}\left(\mathcal{E}+\rho+\mathcal{P}+p\right)=\frac{1}{2}\left(\kappa\dot{m}-nm-\left(\rho+p\right)\right)\,.\label{eq:Friedmann_2}
\end{equation}
The equation of motion for the scalar $\varphi$ is 
\begin{equation}
\dot{m}\, D+3n\left(H-\frac{1}{2}\kappa m\right)+\mathcal{E}_{,\varphi}=\frac{3}{2}\kappa(\rho+p)\,,\label{eq:equation_of_motion_general}
\end{equation}
where 
\begin{equation}
D=\frac{\varepsilon_{,m}}{m}+3H\kappa_{,m}+\frac{3}{2}\kappa^{2}\,.\label{eq:D_general}
\end{equation}
The partial derivatives are all taken at constant Hubble parameter
$H$. The perturbations are not ghosts provided $D>0$. Finally the
sound speed is given by the formula 
\begin{equation}
c_{\text{s}}^{2}=\frac{n+\dot{\kappa}+\kappa\left(H-\kappa m/2\right)}{Dm}\,.\label{eq:Sound_Speed_general}
\end{equation}
Fluids with \emph{Kinetic Gravity Braiding} (including the\emph{ Galileon
}fluid) are conservative---they do not have a notion of entropy---but
they are nonetheless imperfect fluids. One of the peculiarities of
these systems is that the local properties not only depend on the
fluid variables, which in the cosmological context are $\varphi$
and $m$, but also on the expansion (Hubble parameter) and on the
external matter energy density $\rho$ and pressure $p$, see equation
of motion (\ref{eq:equation_of_motion_general}). In particular, the
condition for the absence of ghosts (\ref{eq:D_general}) explicitly
depends on $H$, i.e.~on the external energy density, while the sound
speed not only depends on the Hubble parameter, i.e.~on $\rho$,
but also directly on the external pressure $p$, because of the $\dot{\kappa}$
term in (\ref{eq:Sound_Speed_general}) and the structure of the equation
of motion (\ref{eq:equation_of_motion_general}). Thus, contrary to
$D$, the sound speed also depends on the external equation of state
$w=p/\rho$. This dependence of local properties of fluids with \emph{Kinetic
Gravity Braiding} on external $\rho$ and $p$ reveals a similarity
of these fluids with open systems. 

These properties are crucial for our analysis and the appearance of
the superluminality. 
\section{Robust subluminality without external matter }

\label{sec:RobSub}
\begin{figure}[t]
\begin{centering}
\includegraphics[width=0.97\columnwidth]{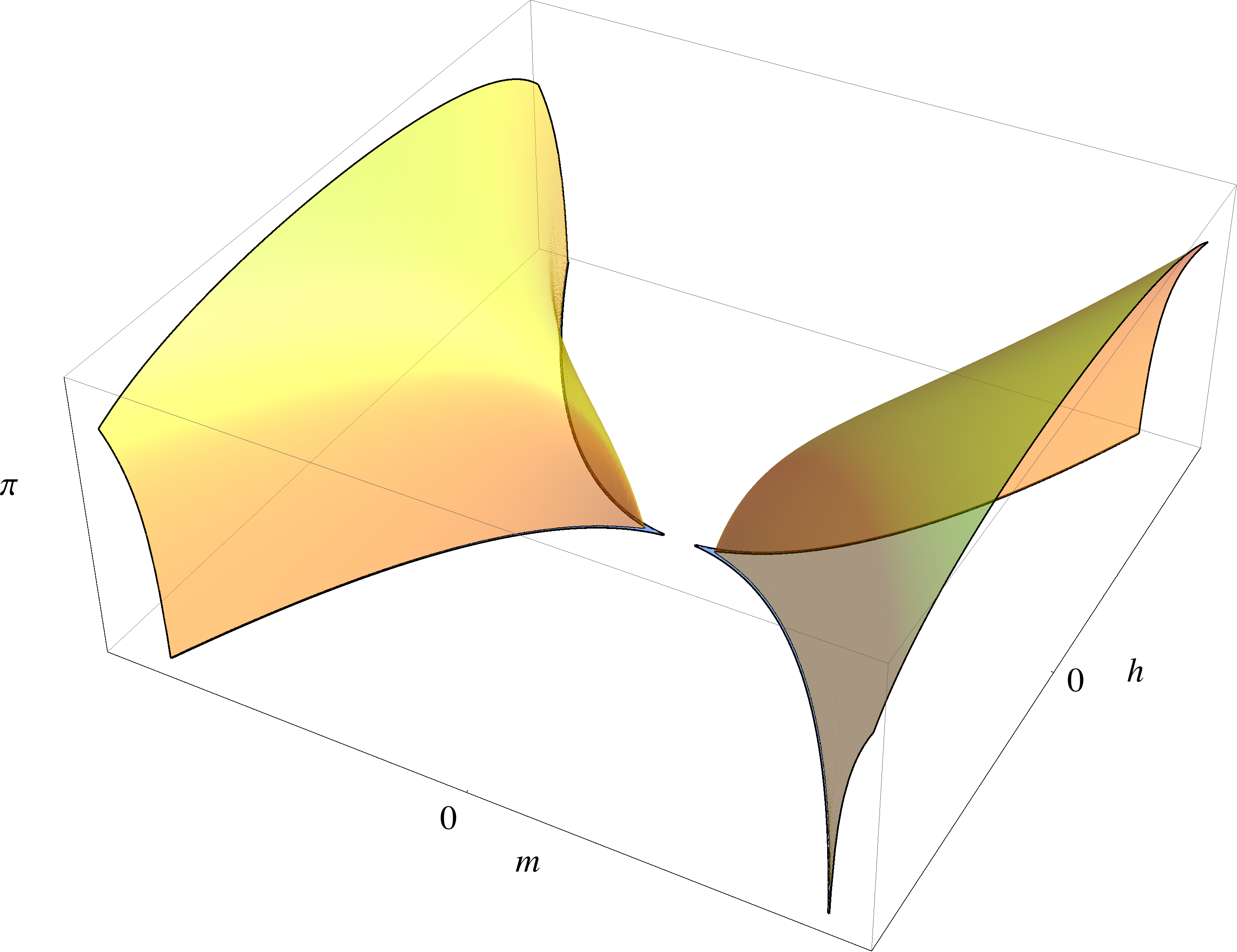}
\par\end{centering}

\caption{\label{fig:NoMatterPhaseSpace}The yellow hypersurface represents the phase space for the cosmological dynamics of the  \emph{Subluminal Galilean
Genesis} given by the system of first order differential equations \eqref{eq:chemical_Pot},  \eqref{eq:Friedmann_2} and \eqref{eq:equation_of_motion_general} with the constraint \eqref{eq:Friedmann_1}. The external energy density and pressure are taken to be zero in all equations. The numerical resolution is  insufficient to plot the origin of the phase space.}
\end{figure}

In this section, we will work with the $\pi$ field
and action (\ref{eq:action}) without any external matter. First of
all, we eliminate $\gamma$ by rescaling (\ref{eq:rescaling}) so
that the time variable which enters both $\dot{\pi}$ and $H$
\begin{equation}
t=\gamma^{1/2}\tau\,,\label{eq:new_time}
\end{equation}
 and consequently the rescaled variables $m$ and $h$ are defined
by 
\begin{eqnarray}
m=\dot{\pi}\,\gamma^{1/2}\,, & \text{and} & h=H\,\gamma^{1/2}\,.\label{eq:rescaled_m_and_h}
\end{eqnarray}
We will denote differentiation with respect to this new time variable
with a prime, $(\ )'\equiv\mbox{d}/\mbox{d}\tau$. Note that we use the same notation, $m$, as in \eqref{eq:chemical_Pot}  for the non-rescaled effective mass of a unit shift-charge.    
The cosmological dynamics are described by the system of three first order differential equations \eqref{eq:chemical_Pot},  \eqref{eq:Friedmann_2} and \eqref{eq:equation_of_motion_general} with the constraint \eqref{eq:Friedmann_1}, where the external energy density and pressure are taken to be zero in all equations. Thus the phase space is a 2d hypersurface in the space $\left(\pi,m,h\right)$. For a similar analysis of a phase-space geometry, see e.g. \cite{Felder:2002jk}.  The phase space cannot be uniquely projected into the plane $\left(\pi,m\right)$, which corresponds to a natural parametrisation of \emph{Galileon} states, see Fig. \ref{fig:NoMatterPhaseSpace}.  Instead, following our paper \cite{Easson:2011zy}, it is useful to describe
dynamics in phase space $(m,h)$ by solving the Friedmann equation
(\ref{eq:Friedmann_1}) with respect to $e^{2\pi}$: 
\begin{equation}
e^{2\pi}=\frac{3m^{4}\beta+12hm^{3}-6h^{2}}{2m^{2}}\,.\label{eq:FriedmannRescaled}
\end{equation}
Thus the expression on the r.h.s. must be strictly positive. Wherever
the above function is negative, it corresponds to a region of the
phase space which is dynamically inaccessible. The advantage of the
variables $(m,h)$ is that one avoids in this way branches in the
square root and related folding of the phase space present for $\left(\pi,m\right)$
description, see discussion in the subsection \ref{sub:PhiMRho}. 

In the variables $(m,h)$ the evolution of the system is given by
equations (\ref{eq:equation_of_motion_general}) and (\ref{eq:Friedmann_2})
which after elimination of $\pi$ and rescaling can be written as
\begin{align}
 & m'=m\frac{\left[12h^{2}-m^{2}\left(m^{2}-1\right)\beta+hm\left(4-6m^{2}+\beta\right)\right]m^{3}-2h^{2}\left(3h+m\right)}{2h^{2}+m^{4}\left(2m^{2}+\beta\right)}\,,\label{eq:EoM_no_matter}\\
 & h'=\frac{\left[12h^{2}m^{2}+\left(8hm^{3}-4h^{2}+m^{4}\beta\right)(1+\beta)+m^{4}\beta\right]m^{4}-12h^{4}}{2\left(2h^{2}+m^{4}\left(2m^{2}+\beta\right)\right)}\,.\nonumber 
\end{align}
We have plotted this phase space for a number of values of $\beta$
in Fig.~\ref{fig:NoMatter}.

This system evolves in the region $\Phi(m,h)$ restricted because
of (\ref{eq:FriedmannRescaled}) to: 
\begin{equation}
m^{4}\beta+4hm^{3}-2h^{2}>0\,.\label{eq:region_of_motion}
\end{equation}
Thus the phase space $\Phi(m,h)$ is located between two curves 
\begin{equation}
h_{\pm}\left(m\right)=m^{2}\left[m\pm\sqrt{m^{2}+\frac{\beta}{2}}\,\right]\,.\label{eq:bordersOfSpace}
\end{equation}
Further, the absence of ghosts requires the positiveness of $D$ from
(\ref{eq:D_general}) and provides the bound 
\begin{equation}
\frac{m^{2}}{3}D=2h^{2}+m^{4}\left(2m^{2}+\beta\right)>0\,,\label{eq:D_No_matter}
\end{equation}
which is always satisfied for $\beta=1+\alpha>1$. On Fig.~\ref{fig:NoMatter},
we have marked the inaccessible regions in deep red with the boundaries
defined by (\ref{eq:bordersOfSpace}). There are no ghosty regions.

The general formula (\ref{eq:Sound_Speed_general}) for the sound
speed translates for $c_{\text{s}}^{2}\left(m,h\right)$ to 
\begin{equation}
c_{\text{s}}^{2}=1-\frac{12}{\left(Dm\right)^{2}}\,\, m\mathscr{P}\left(m,h\right)\,,\label{eq:sound_speed_no_matter}
\end{equation}
where the polynomial cubic in $h$ 
\begin{equation}
\mathscr{P}\left(m,h\right)=8h^{3}+2h^{2}m\left(1-4m^{2}+\beta\right)+4hm^{4}\left(2m^{2}-1\right)+m^{5}\left(4m^{4}+5m^{2}\beta+(\beta-1)\beta\right)\,,\label{Polynom}
\end{equation}
defines whether the sound speed is less or larger than the speed of
light. Subluminality corresponds to $m\mathscr{P}\left(m,h\right)>0$.
For each $m$, the polynomial $\mathscr{P}\left(m,h\right)$ can have
maximum 3 real roots $h_{\text{s}}\left(m\right)$. The number of
roots, $ $when solved for $h$, is defined by the sign of the discriminant
$\mathscr{D}\left(\alpha,m\right)$, 
\begin{align}
 & -\frac{\mathscr{D}\left(\alpha\right)}{32m^{8}}=2144\, m^{10}+16\, m^{8}\,\left[125+203\alpha\right]+2\, m^{6}\,\left[339+\alpha(1066+915\alpha)\right]+\label{eq:Discriminant_NO_matter}\\
 & +4\, m^{4}\,\left[38+\alpha(126+\alpha(194+115\alpha))\right]+m^{2}\,\left[32+\alpha(116+\alpha(154+\alpha(119+47\alpha)))\right]+\nonumber \\
 & +\alpha(1+\alpha)(2+\alpha)^{3}\,.\nonumber 
\end{align}
This is the only formula for which the parameter $\alpha=\beta-1$
is more convenient. The polynomial on the r.h.s. has only even powers
of $m$ and is manifestly positive for all non-negative $\alpha$.
Thus the discriminant $\mathscr{D}\left(\alpha,m\right)$ of the polynomial
$\mathscr{P}\left(m,h\right)$ is always negative for all $m\neq0$
and non-negative $\alpha$, whereas $\mathscr{D}\left(\alpha,0\right)=0$.
Therefore, for $m\neq0$, the polynomial $\mathscr{P}\left(m,h\right)$
has only one real root $h_{\text{s}}\left(m\right)$ which one can
find analytically, but the expression for which is by far too long
to be suitable for any analysis %
\footnote{Note that for $\alpha<0$ this situation changes so that for sufficiently
small $m$ the polynomial $\mathscr{P}\left(m,h\right)$ has 3 real
roots and there is always a superluminal region, see the lower right
panel in Fig. \ref{fig:NoMatter}. %
}. Now we can evaluate the value of $m\mathscr{P}\left(m,\, h_{\pm}\left(m\right)\right)$
at the borders $h_{\pm}\left(m\right)$ of the phase space $\Phi(m,h)$
given by (\ref{eq:bordersOfSpace}). For $h_{-}\left(m\right)$ we
obtain 
\begin{equation}
m\mathscr{P}\left(m,\, h_{-}\left(m\right)\right)=\frac{8m^{6}}{\left(2m^{2}+\beta\right)^{2}}\left(\xi^{4}-4\xi^{3}+9\xi^{2}-8\xi+4\right)\,,\label{eq:Border_}
\end{equation}
where $\xi=m/\sqrt{m^{2}+\beta/2}$. This function $m\mathscr{P}\left(m,\, h_{-}\left(m\right)\right)$
is always non-negative since the polynomial in $\xi$ is positive
for $\xi=0$ and has four complex roots $\frac{1}{2}\left(2-i\pm\sqrt{-5-4i}\right)$
and the complex conjugates. For the other boundary, $h_{+}\left(m\right)$,
we have 
\begin{equation}
m\mathscr{P}\left(m,\, h_{+}\left(m\right)\right)=\frac{8m^{6}}{\left(2m^{2}+\beta\right)^{2}}\left(\xi^{4}+4\xi^{3}+9\xi^{2}+8\xi+4\right)\,,\label{eq:Border+}
\end{equation}
which is also always non-negative, because the polynomial in $\xi$
has four complex roots of the form $\frac{1}{2}\left(-2-i\pm\sqrt{-5-4i}\right)$
and the complex conjugates. 

This means that the single real root $h_{\text{s}}\left(m\right)$
of $\mathscr{P}\left(m,h\right)$ is not located in the phase space
and $m\mathscr{P}\left(m,h\right)\geq0$ not only on the boundaries
$h_{\pm}\left(m\right)$ but the whole cosmological phase space $\Phi(m,h)$.
Therefore, for all $\alpha\geq0$, we conclude that the system is
\emph{subluminal} on all cosmological configurations. In this sense
we have proven a strong version of the desired property 6 on page
2 from \cite{Creminelli:2012my}. 

We have illustrated this in Fig.~\ref{fig:NoMatter}. The roots $h_{\text{s}}(m)$
of (\ref{Polynom}) are the boundaries of the blue regions. As can
be seen for $\beta\geq1$ ($\alpha\geq0$) there is only one such
boundary and it always occurs inside the deep red dynamically inaccessible
region, as we have proven. Only in the case of $\beta=0$ ($\alpha=-1$),
which is not part of the parameter space of \emph{Subluminal Galilean
Genesis, }a new region of superluminality appears: now, for a range
of values of $m$, there are three real roots $h_{\text{s}}(m)$ delineating
the boundaries of these superluminal regions. Two of these roots lie
inside the phase space.

\section{Superluminality with external matter}

\label{sec:superlu}In this section we consider cosmology with the
\emph{Galileon} in the presence of external matter with a standard
equation of state $w=p/\rho=\mbox{const}$, e.g.~dust or radiation.
We will restrict our attention to matter satisfying the Strong Energy
Condition (SEC) and the Dominant Energy Condition (DEC) so that $-1/3\leq w\leq1$.
The main purpose of this section is to prove that when such external
mater is added into the game there are some regions in cosmological
phase space where the fluctuations of the \emph{Galileon} propagate
faster than light. Moreover, in some of these regions the perturbations
are not ghosty so that these configurations should belong to the same
effective filed theory (EFT) which describes the regions where the
perturbations are subluminal and those where the external matter is
negligible or absent at all.

\subsection{$\left(m,h,\rho\right)$ coordinates and a proof of general superluminality }

\label{sub:super-mh}First, similarly to the previous section, we
can proceed by working in the phase space $\left(m,h,\rho\right)$
where $\rho$ is rescaled as in (\ref{eq:Rho_rescaled}). One can
eliminate $\pi$ by solving the first Friedmann equation (\ref{eq:Friedmann_1})
with respect to $\pi$ so that 
\[
e^{2\pi}=\frac{2\rho+3m^{4}\beta+12hm^{3}-6h^{2}}{2m^{2}}\,.
\]
Therefore the phase space is restricted by 
\begin{equation}
\frac{2}{3}\rho-2h^{2}+4hm^{3}+m^{4}\beta>0\,.\label{eq:Border_hm_With_Rho}
\end{equation}
The perturbations of the \emph{Galileon} are not ghosts when $D$
from (\ref{eq:D_general}) is positive which provides the inequality
\begin{equation}
m^{2}D=6h^{2}+6m^{6}+3m^{4}\beta-2\rho>0\,.\label{eq:D_Rho}
\end{equation}
We would like to stress that, contrary to k-\emph{essence, }the coefficient
$D$ explicitly depends on the external energy density $\rho$. Thus
a sufficiently large positive amount of the external energy density
can turn the perturbations of the \emph{Galileon} into ghosts. Moreover,
in the presence of $\rho$ there are always ghosts close enough to
the origin of the phase space, $m=h=0$.

From the above inequalities, it follows that the external energy density
should be located 
\begin{equation}
3h^{2}+\frac{3}{2}m^{4}\beta-6hm^{3}<\rho<3h^{2}+\frac{3}{2}m^{4}\beta+3m^{6}\,.\label{eq:borders_of_rho}
\end{equation}
This is only possible for such $\left(m,h\right)$ that 
\begin{equation}
m\left(m^{3}+2h\right)>0\,.\label{eq:borders_Of_mh_for_reasonable_Rho}
\end{equation}
In particular this works for $m>0$ and $h>0$. For the sound speed
(\ref{eq:Sound_Speed_general}) one obtains 
\begin{equation}
c_{\text{s}}^{2}=1-\frac{4}{\left(Dm\right)^{2}}\,\, m\left[3\mathscr{P}\left(m,h\right)-\rho\left(8h+2m\left(\beta+1\right)+m^{3}\left(1+3w\right)\right)\right]\,,\label{eq:sound_with_Rho}
\end{equation}
where $\mathscr{P}\left(m,h\right)$ is given by (\ref{Polynom})
and is the same polynomial as in the previous section. For $m>0$
and $8h+2m\left(\beta+1\right)+m^{3}\left(1+3w\right)>0$, the sound
speed is larger than the speed of light provided 
\begin{equation}
\rho>\rho_{\text{sup}}=\frac{3\mathscr{P}\left(m,h\right)}{8h+2m\left(1+\beta\right)+m^{3}\left(1+3w\right)}\,.
\end{equation}
If, contrary to the matter-free case, dynamics (\ref{eq:Border_hm_With_Rho})
allow for such $\left(m,h\right)$ that $m\mathscr{P}\left(m,h\right)<0$,
then a positive amount of external matter which satisfies the no-ghost
condition (\ref{eq:D_Rho}) will make the perturbations superluminal.
\\

Now let us prove that it is always possible to find such $\rho$.
It is convenient to introduce a ``probe'' energy density 
\begin{equation}
\rho_{\lambda}\left(m,h\right)=3h^{2}+\frac{3}{2}m^{4}\beta+\lambda m^{6}\,,\label{eq:Probe}
\end{equation}
We restrict the numerical parameter $\lambda$ to $0<\lambda<3$.
This restriction ensures that probe energy density is always positive
and fits into the borders (\ref{eq:borders_of_rho}) for $m>0$ and
$h>0$. On this probe energy density we have 
\begin{equation}
c_{\text{s}}^{2}=1+\frac{Z\left(m,h\right)}{2m^{6}(\lambda-3)^{2}}\,,\label{eq:SoundProbe}
\end{equation}
where 
\begin{align}
 & Z\left(m,h\right)=18h^{2}(3+w)+8hm\left[3(1+\beta)+2m^{2}(\lambda-3)\right]+\label{eq:ZonProbe}\\
 & +m^{2}\left[12\beta+m^{2}(9(w-3)\beta+4(1+\beta)\lambda)+2m^{4}(\lambda\left(1+3w\right)-12)\right]\,.\nonumber 
\end{align}
$\lambda=3$ is the boundary of the ghosty region on which the sound
speed (\ref{eq:sound_with_Rho}) diverges. For normal matter with
$w>-3$, the polynomial $Z$ as a function of $h$ is a convex parabola.
Thus for sufficiently large $h$ this function $Z$ is always positive.
On the other hand, for given $\left(m,h\right)$, we can choose $\lambda$
arbitrarily close to $3$, i.e.~arbitrarily close to the ghosty region.
Thus we conclude that the sound speed of the \emph{Galileon} can not
only be superluminal but can, indeed, acquire arbitrarily large positive
values. We would like to stress that by construction the perturbations
are not ghosts on these configurations. This conclusion only uses
the positivity of $\beta$ which is needed for the positivity of $\rho_{\lambda}$.
\\

One may think that the superluminality can only correspond to large
$h$ but in fact there are always regions around $h=0$ where the
sound speed is larger than 1. Indeed, in that case 
\begin{equation}
\frac{3}{2}m^{4}\beta<\rho<\rho_{\text{gh}}=\frac{3}{2}m^{4}\beta+3m^{6}\,,\label{eq:Rho_Limit_HZero}
\end{equation}
and 
\begin{equation}
c_{\text{s}}^{2}=1-\frac{4m^{4}\left[3m^{4}\left(4m^{4}+5m^{2}\beta+(\beta-1)\beta\right)-\rho\left(2\left(\beta+1\right)+m^{2}\left(1+3w\right)\right)\right]}{\left(6m^{6}+3m^{4}\beta-2\rho\right)^{2}}\,.\label{eq:SoundSpeed_HZero}
\end{equation}
Thus, for $w>-1/3$ which are considering here, one has to require
that 
\begin{equation}
\rho>\rho_{\text{sup}}=\frac{3m^{4}\left(4m^{4}+5m^{2}\beta+(\beta-1)\beta\right)}{2\left(\beta+1\right)+m^{2}\left(1+3w\right)}\,,
\end{equation}
and there are $\rho$ realising this regime provided $\rho_{\text{sup}}<\rho_{\text{gh}}$.
This happens for those $m$ for which 
\begin{equation}
6m^{4}(w-1)+m^{2}\left(4-\beta(5-3w)\right)+4\beta>0\,.
\end{equation}
The latter inequality is manifestly satisfied for sufficiently small
$m^{2}$ because $\beta\geq1$. Thus the superluminality without ghosts
can also happen for small $m$ and $h$ or in other words for $\rho\ll1$. 

\selectlanguage{english}%

\subsection{\textmd{\normalsize $\left(\phi,m,\rho\right)$}\foreignlanguage{british}{
coordinates and induced superluminality of original configurations }}

\selectlanguage{british}%
\label{sub:PhiMRho}The disadvantage of choosing the phase space in
terms of $(m,h,\rho)$ in section \ref{sub:super-mh} is that adding
external energy density $\rho$ while keeping both $m$ and $h$ fixed
forces the value of the scalar field $\pi$ to change. Thus, in some
sense, the new configuration including $\rho$ does not correspond
to the initial one. To avoid this, in this section, we will consider
the phase space in terms of the coordinates $\left(\phi,m,\rho\right)$,
where $\phi$ is given by (\ref{eq:new_field_phi}), $m=\pi'=\phi'/\phi$,
i.e.~is still given by (\ref{eq:rescaled_m_and_h}) and $\rho$ is
again rescaled as in (\ref{eq:Rho_rescaled}). However, we will see
this is much more complicated as a result of the fact that this three-dimensional
phase space is double folded, because of the square root 
\begin{equation}
\Omega_{\rho}=\sqrt{6\left[2\rho+m^{2}\left(3\beta m^{2}+6m^{4}-2\phi^{2}\right)\right]}\,,\label{eq:definiton_of_Omega}
\end{equation}
in the solution of the Friedmann equation (\ref{eq:Friedmann_1})
\begin{equation}
h=m^{3}+\frac{1}{6}\Omega_{\rho}\,.\label{eq:solution_of_Friedmann_with_matter}
\end{equation}
Here $\Omega_{\rho}=\Omega\left(\phi,m,\rho\right)$ is defined in
such a way that the sign of $\Omega_{\rho}$ changes when the argument
of the square root evolves trough zero. Only regions where the argument
of the square root in $\Omega_{\rho}$ is positive $ $correspond
to physically available configurations, which result in a condition
on $\rho,$ 
\begin{equation}
\rho\geq\phi^{2}-\frac{3}{2}m^{2}\left(\beta+2m^{2}\right)\,.\label{eq:bound_on_rho_from_Omega}
\end{equation}

In this discussion, we are only considering normal matter and parameter
ranges relevant to the \emph{subluminal Galilean Genesis} model, i.e.
we will assume in what follows that 
\begin{align}
-\nicefrac{1}{3} & \leq w\leq1\,,\quad w=\text{const}\label{eq:Props}\\
1 & \leq\beta\leq4\,.\nonumber 
\end{align}

Let us introduce the configurations of the \emph{Galileon} which are
dynamically allowed \emph{without any external matter}:
\begin{eqnarray}
\Gamma=\left(\phi,m\right): & \text{such configurations that} & \phi^{2}\leq\frac{3}{2}m^{2}\left(\beta+2m^{2}\right)\,.\label{eq:Original_Configurations}
\end{eqnarray}
These configurations are exactly those which we have studied in the
section \ref{sec:RobSub}, just in different variables. In the rest
of this section we will only consider these configurations $\Gamma$.
If we add any positive $\rho\geq0$ to any of these configurations
$\Gamma$, it is still dynamically allowed, since (\ref{eq:bound_on_rho_from_Omega})
holds. The addition of the positive energy density $\rho$ opens up
the phase space---allows one to probe new cosmological configurations
$\left(\phi,m\right)$ violating the condition (\ref{eq:Original_Configurations}). 

For the positivity of $\rho$ we have to require that 
\begin{equation}
\text{\ensuremath{\rho}\ positive:\qquad}\Omega_{\rho}^{2}>\Omega_{0}^{2}=6m^{2}\left[3\beta m^{2}+6m^{4}-2\phi^{2}\right]\,.\label{eq:Omega_Zero}
\end{equation}
On configurations $\Gamma$ we always have $\Omega_{0}^{2}>0$. \\
\\
As we have shown before, in the section \ref{sec:RobSub} all configurations
$\Gamma$ have subluminal sound speed when there is no external matter.
In the subsection \ref{sub:super-mh} we have showed that superluminality
takes place somewhere on available and not ghosty phase space. However,
these regions could not be available without external matter -- so
that for $\left(\phi,m\right)$ the condition (\ref{eq:Original_Configurations})
is violated. Our main goal now is to prove that around some of the
configurations $\Gamma$ the sound speed can become superluminal on
the addition of external matter, but the perturbations remain not
ghosty. 

The perturbations are not ghosts when $D$ from (\ref{eq:D_general})
is positive, which provides in our current variables the inequality
\begin{eqnarray}
\frac{\phi^{2}D}{2}=m\left(\Omega_{\rho}-\Omega_{\text{ghosts}}\right)>0\,, & \:\text{where}\: & \:\Omega_{\text{ghosts}}\equiv\frac{\phi^{2}-3m^{2}\left(\beta+3m^{2}\right)}{m}\,.\label{eq:No_Ghost_region}
\end{eqnarray}
Therefore the condition for the absence of ghosts reduces to
\begin{eqnarray}
\text{No ghosts:\qquad\ }\Omega_{\rho}>\Omega_{\text{ghosts}}\,, & \text{for} & m>0\,,\nonumber \\
\Omega_{\rho}<\Omega_{\text{ghosts}}\,, & \text{for} & m<0\,,\label{eq:No_Ghosts_Omega_Conditions}
\end{eqnarray}
Note that $\Omega_{\text{ghosts}}\left(\Gamma\right)>0$ for $m<0$
and $\Omega_{\text{ghosts}}\left(\Gamma\right)<0$ for $m>0$. Further
it is convenient to express 
\begin{equation}
\Omega_{\text{ghosts}}=-\frac{\Omega_{0}^{2}+18m^{4}\left(\beta+4m^{2}\right)}{12m^{3}}\,.\label{eq:OmegaGhost_vs_Omega_Zero}
\end{equation}
Hence, on $\Gamma$ and for sufficiently small $m^{2}$ one always
obtains $\Omega_{\text{ghosts}}^{2}\left(\Gamma\right)>\Omega_{0}^{2}\left(\Gamma\right)$.
\\

The sound speed (\ref{eq:Sound_Speed_general}) as a function on phase
space $\left(\phi,m,\rho\right)$ is given by 
\begin{equation}
c_{\text{s}}^{2}\left(\phi,m,\rho\right)=1-\frac{m\mathscr{R}_{w}\left(\phi,m,\Omega_{\rho}\right)}{3\phi^{4}D^{2}}\,,\label{eq:sound_speed_with_matter}
\end{equation}
where 
\begin{align}
 & \mathscr{R}_{w}\left(\phi,m,\Omega_{\rho}\right)=(7-3w)\Omega_{\rho}^{2}m+16\Omega_{\rho}\left(3m^{2}\left(\beta+3m^{2}\right)-\phi^{2}\right)+\label{eq:polynomial_defining_super}\\
 & +6m\left[3m^{2}\left(4\beta^{2}+3(9+w)\beta m^{2}+6(7+w)m^{4}\right)-2\phi^{2}\left(2(1+\beta)+3(3+w)\text{ }m^{2}\right)\right]\,.\nonumber 
\end{align}
The superluminality is present in regions where $m\mathscr{R}_{w}\left(\phi,m,\Omega_{\rho}\right)<0$.
$\mathscr{R}_{w}\left(\phi,m,\Omega_{\rho}\right)$ depends on $\rho$
only through $\Omega_{\rho}$, with respect to which $m\mathscr{R}_{w}\left(\phi,m,\Omega_{\rho}\right)$
is a convex parabola. Thus superluminality is only possible if the
roots $\Omega_{w\pm}$ of $\mathscr{R}_{w}\left(\phi,m,\Omega_{\rho}\right)$
are real. Given that, in order to have superluminality, $\Omega_{\rho}$
must lie between these real roots: 
\begin{align}
\text{Superluminality exists:\qquad}\Omega_{w-} & <\Omega_{\rho}<\Omega_{w+\,,}\label{eq:Omega_Root_General}\\
\Omega_{w\pm} & =\frac{8}{7-3w}\left[\Omega_{\text{ghosts}}\pm\sqrt{\Omega_{\text{ghosts}}^{2}+\Sigma}\right]\,,\ \Omega_{w\pm}\in\mathbb{R}\,,\nonumber 
\end{align}
where $\Omega_{\text{ghosts}}$ is defined in (\ref{eq:No_Ghost_region})
and 
\begin{equation}
\Sigma=-\frac{(7-3w)}{16}\left[9m^{2}\left[(\beta-1)\beta+(7\beta-2)m^{2}+12m^{4}\right]+\left[2(1+\beta)+3(3+w)m^{2}\right]\frac{\Omega_{0}^{2}}{4m^{2}}\right]\,,\label{eq:SigmaDeff}
\end{equation}
with $\Omega_{0}$ given by (\ref{eq:Omega_Zero}). Now we will prove
that there are such $\left(\phi,m\right)$ from $\Gamma$ that these
roots exist and that at least some $\Omega_{\rho}$ between these
roots do not have ghosts and have $\rho>0$. 

For $\Omega_{w\pm}$ of (\ref{eq:Omega_Root_General}) to be real,
we need to require that 
\begin{equation}
\Omega_{\text{ghosts}}^{2}+\Sigma\geq0\,.\label{eq:existence}
\end{equation}
For our chosen parameters $w$ and $\beta$, (\ref{eq:Props}), $\Sigma\left(\Gamma\right)<0,$
on \emph{all }configurations $\Gamma$. Therefore, if $\Omega_{w\pm}$
exist somewhere on $\Gamma$, then for $m>0$ we have $\Omega_{w-}\left(\Gamma\right)\leq\Omega_{w+}\left(\Gamma\right)<0$
whereas for $m<0$ these roots are located $0<\Omega_{w-}\left(\Gamma\right)\leq\Omega_{w+}\left(\Gamma\right)$.

To avoid ghosts as in (\ref{eq:No_Ghosts_Omega_Conditions}), we have
to require that
\begin{align}
\Omega_{w+}= & \frac{8}{7-3w}\left[\Omega_{\text{ghosts}}+\sqrt{\Omega_{\text{ghosts}}^{2}+\Sigma}\right]>\Omega_{\text{ghosts}}\,,\quad\text{for }m>0\,,\label{eq:OmegaPlusLarger}\\
\Omega_{w-}= & \frac{8}{7-3w}\left[\Omega_{\text{ghosts}}-\sqrt{\Omega_{\text{ghosts}}^{2}+\Sigma}\right]<\Omega_{\text{ghosts}}\,,\quad\text{for }m<0\,.\nonumber 
\end{align}
For configurations $\Gamma$ and normal matter these are equivalent
to the condition 
\begin{equation}
\text{Superluminal \& no ghosts:\quad}\Sigma>\Omega_{\text{ghosts}}^{2}\left(\left(\frac{1+3w}{8}\right)^{2}-1\right)\,,\label{eq:Superluminal_without_ghosts_forGamma}
\end{equation}
which is a stronger condition than the condition (\ref{eq:existence})
for the existence of roots $\Omega_{w\pm}$ and which implies 
\begin{equation}
\Omega_{w-}<\Omega_{\text{ghosts}}<\Omega_{w+}\,.
\end{equation}

Therefore if (\ref{eq:Superluminal_without_ghosts_forGamma}) holds
and $m^{2}$ is sufficiently small, one can always adjust $\Omega_{\rho}$
to be sufficiently close to $\Omega_{\text{ghosts}}$ so that $\Omega_{0}^{2}<\Omega_{\rho}^{2}<\Omega_{\text{ghosts}}^{2}$
and therefore have superluminality with $\rho>0$ and no ghosts.\\

Now let us find simple \emph{sufficient} conditions on $\left(\phi,m\right)$
from $\Gamma$ to satisfy (\ref{eq:Superluminal_without_ghosts_forGamma}),
proving that there are accessible superluminal configurations. 

The inequality (\ref{eq:Superluminal_without_ghosts_forGamma}) is
equivalent to 
\begin{equation}
\mathcal{F}\left(\phi^{2}\right)=\left(3+w\right)\phi^{4}+B\,\phi^{2}+C>0\,,\label{eq:parabola}
\end{equation}
where
\begin{equation}
B=2m^{2}\left[4-\left(5+3w\right)\beta-3m^{2}(3+w)\right]\,,
\end{equation}
and
\begin{equation}
C=3m^{4}\left[3m^{4}(5w-1)+12m^{2}w\beta+(1+3w)\beta^{2}\right]\,.
\end{equation}
If for some range of $m$ there are no real roots for the equation
$\mathcal{F}\left(\phi^{2}\right)=0$, then the inequality (\ref{eq:parabola})
(or equivalently (\ref{eq:Superluminal_without_ghosts_forGamma}))
holds for \emph{all} $\phi$ and in particular for such $\phi$ that
the matter-free configuration belongs to $\Gamma$, (\ref{eq:Original_Configurations}).
Inequality (\ref{eq:parabola}) holds for all $\phi$, provided the
quadratic equation for $\phi^{2}$ does not have positive solutions,
in particular, if the discriminant is negative, i.e. if 
\begin{equation}
S(m^{2})=18m^{4}(1-w)(3+w)-3m^{2}(3+w)(4+\left(3w-5\right)\beta)-4(5+3w-2\beta)\beta+8<0\,.\label{eq:S<0}
\end{equation}
 Let us consider sufficiently small $m^{2}$. For any $\beta\geq1$
and 
\begin{equation}
w>w_{\beta}=\frac{2\left(\beta+\beta^{-1}\right)-5}{3}\,,\label{eq:limiting_W}
\end{equation}
we obtain $S\left(0\right)<0,$ so that for sufficiently small $m^{2}$
we have $S\left(m^{2}\right)<0$. In particular for $\beta=1$ we
have $w_{1}=-\nicefrac{1}{3}$ thus any SEC satisfying matter creates
a small region close to $m=0$ such that the sound speed is superluminal
and yet the perturbations are not ghosts. For $\beta=2$ this translates
into $w_{2}=0$ so that reasonable matter would always create a superluminal
region. Whereas, for $\beta=4$, which is the maximal interesting
value of this parameter from \cite{Creminelli:2012my} we have $w_{4}=\nicefrac{7}{6}$
which would violate the DEC and which is not an equation of state
available for standard matter. The limiting equation of state $w_{\beta}$
corresponds to radiation when 
\begin{equation}
\beta_{\text{rad}}=\frac{3+\sqrt{5}}{2}\simeq2.62\,,\label{eq:Betta_Rad}
\end{equation}
and to ultra-stiff equation of state $w_{\beta}=1$ when 
\begin{equation}
\beta_{\text{stiff}}=2+\sqrt{3}\simeq3.73\,.\label{eq:betta_Stiff}
\end{equation}
Thus for $1\leq\beta\leq\beta_{\text{stiff}}$ the addition of standard
matter creates a superluminal but not ghosty region at least for those
configurations from $\Gamma$ which have sufficiently small $m$.
\\

Now we can look for more general larger $m$. Since $S(m^{2})$ of
(\ref{eq:parabola}) is a convex parabola, if we require that $S(m^{2})=0$
has at least one positive root $m^{2}$, then there will be a range
of $m$ where the inequality (\ref{eq:S<0}) is satisfied and where
consequently (\ref{eq:Superluminal_without_ghosts_forGamma}) holds
for \emph{all} $\phi$. These roots are given by 
\begin{equation}
m_{\pm}^{2}=\frac{12-15\beta+w(4+(4+3w)\beta)\pm\sqrt{(3+w)Y\left(w,\beta\right)}}{12(1-w)(3+w)}\,,
\end{equation}
where 
\begin{equation}
Y\left(w,\beta\right)=(1+w)(11+3w(3w-4))\beta^{2}-8(1+w)(9w-5)\beta+80w-16\,,\label{eq:Y}
\end{equation}
 which is a quadratic polynomial in $\beta$ and cubic in $w$. We
now must ensure that both $Y>0$ and $m_{+}^{2}>0$. For normal matter
$(1+w)(11+3w(3w-4))>0$, thus for $Y$ to be positive, the parameter
$\beta$ must be larger than 
\begin{equation}
\beta_{*}\left(w\right)=\frac{4\left[w(4+9w)+2\sqrt{3(3+w)(1-w^{2})(1-3w)}-5\right]}{(1+w)(11+3w(3w-4))}\,,
\end{equation}
for normal matter with $w<\nicefrac{1}{3}$, and $\beta$ can be arbitrary
for $w>\nicefrac{1}{3}$. It is easy to check that $\beta_{*}<1$
for all $w$ we consider, thus $Y$ is always positive for our set
of parameters (\ref{eq:Props}).

For $0\leq w\leq\nicefrac{1}{3}$ one obtains that $m_{-}^{2}$ is
always negative whereas $m_{+}^{2}$ is positive for either
\begin{eqnarray}
1\leq\beta\leq2 & \text{and} & 0\leq w\leq\nicefrac{1}{3}\,,
\end{eqnarray}
or for
\begin{eqnarray}
2\leq\beta<\beta_{\text{rad}} & \text{and} & \frac{(\beta-2)(2\beta-1)}{3\beta}<w\leq\frac{1}{3}\,.
\end{eqnarray}
 If we allow for matter with an ultra-relativistic equation of state
$1/3\leq w\leq1$ then 
\begin{eqnarray}
\beta_{\text{rad}}\leq\beta<\beta_{\text{stiff}} & \text{and} & \frac{(\beta-2)(2\beta-1)}{3\beta}<w<1\,.
\end{eqnarray}
Thus we have proved that for any model with $\beta<\beta_{\text{stiff}}\simeq3.73$
one can add a positive amount of external energy energy density with
$0<w<1$ to make the sound speed of the perturbations around at least
some original configurations larger than the speed of light. We presume
that this statement also holds for $\beta_{\text{stiff}}<\beta<4$.
But to prove this it would require a more cumbersome analysis e.g.
requiring the roots $\phi^{2}$ of $\mathcal{F}\left(\phi^{2}\right)=0$
to be negative. We should also stress that we have only constructed
one type of configurations with superluminality. We have not performed
an exhaustive search to classify all such configurations.\\
\section{Results Summary and Discussion}
\begin{figure}[t]
\begin{centering}
\includegraphics[width=0.97\columnwidth]{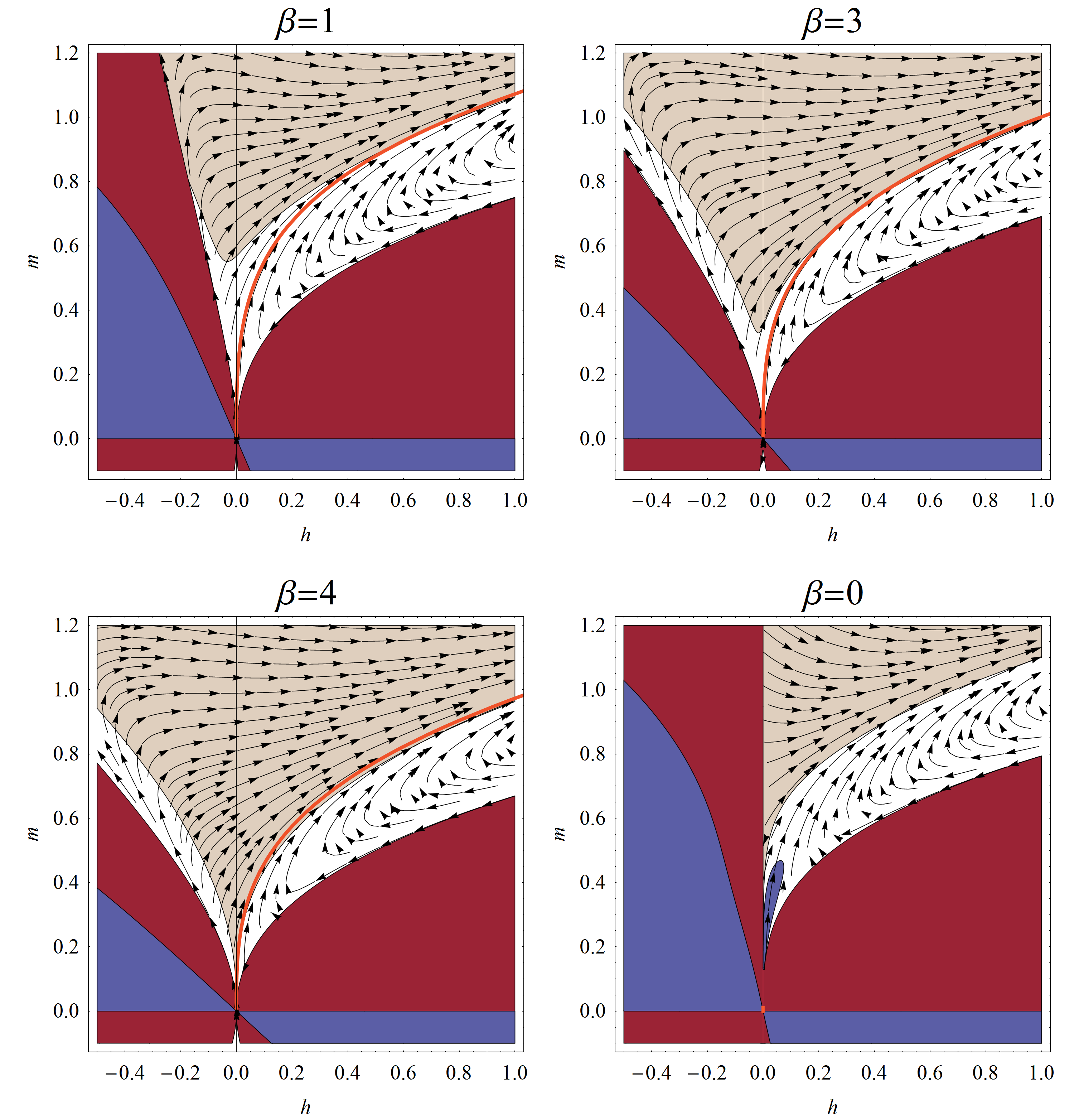}
\par\end{centering}

\caption{\label{fig:NoMatter}Phase portrait for the \emph{Subluminal Galilean
Genesis }system (\ref{eq:EoM_no_matter}) for various values of the
parameter $\beta\equiv1+\alpha$ with no external matter present.
The axes are rescaled coordinates $h=(f/\Lambda)^{3/2}H$, $m=(f/\mbox{\ensuremath{\Lambda}})^{3/2}\dot{\pi}$.
The deep red region is dynamically inaccessible, the blue corresponds
to configurations where the sound speed is superluminal, the sandy
region---to those configurations where $c_{\text{s}}^{2}<0$ and gradient
instabilities are present. Healthy trajectories evolve through the
white regions and have been illustrated with flow lines. The \emph{Galilean
Genesis} trajectory discussed in \cite{Creminelli:2010ba,Creminelli:2012my}
is a separatrix and has been explicitly singled out and marked as
the thick salmon-coloured line. Nowhere in this phase space is there
a region where the perturbations are ghosty.\protect \\
\protect \\
For $\beta\geq1$, the accessible phase space \emph{never }has any
superluminality \emph{anywhere}. This includes the original \emph{Galilean
Genesis }model with $\beta=1$. As $\beta$ increases, the sound speeds
are reduced and eventually for $\beta=4$ the sandy region with imaginary
sound speed reaches to the origin of the phase space which should
correspond to a singular Minkowski space. For $\beta<1$, an accessible
superluminal region appears at the origin and grows in extent with
decreasing $\beta$. \protect \\
}
\end{figure}

In the case when there is no external matter, we have analytically proved in section \ref{sec:RobSub} that for a spatially flat Friedmann universe the perturbations of the \emph{Galileon} are never superluminal. 
To illustrate our results we have plotted  phase flows in coordinates $(m,h)$ (which are rescaled $(\dot\pi,H)$) in Fig.~\ref{fig:NoMatter}.
The red regions are dynamically inaccessible as the r.h.s.~of \eqref{eq:FriedmannRescaled} is negative, but there are no ghosty
regions in the phase space at all. The trajectories correspond to solutions of \eqref{eq:EoM_no_matter}. The \emph{Galilean Genesis }trajectory
is marked as the thick salmon line originating at the origin. The
regions of superluminality are coloured in deep blue. We can clearly
see that for $\beta\geq1$ they never occur in the phase space. Indeed
there is no superluminality anywhere in the phase space even in the
original \emph{Galilean Genesis }scenario of \cite{Creminelli:2010ba}.

\begin{figure}[tb]
\begin{centering}
\begin{overpic}[width=15cm]{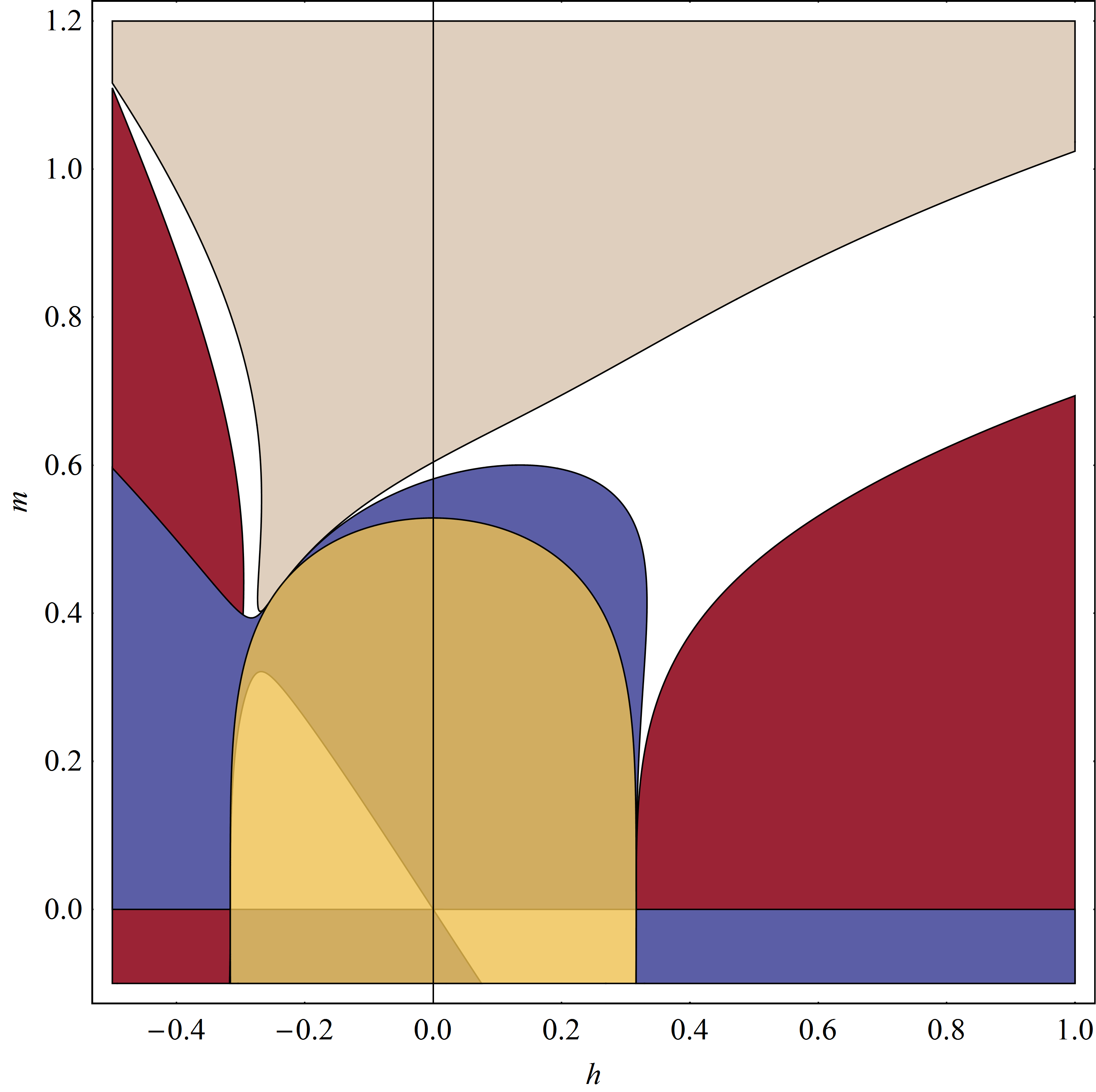}
\put(67,35){\Huge Dynamically}
\put(67.3,29){\Huge inaccessible}
\put(44,80){\Huge $c_\text{s}^2<0$}
\put(57,60){\Huge $0<c_\text{s}^2<1$}
\put(32.5,20){\Huge Ghosts}
\put(43,54.5){\begin{turn}{-30} {\large Superluminal}\end{turn}}
\end{overpic}
\par\end{centering}

\caption{\label{fig:withrho}A slice through the phase space of the \emph{Subluminal
Galilean Genesis }model with $\beta=2$ in the presence of external
radiation ($w=\nicefrac{1}{3}$) with rescaled energy density $\rho=0.3$.
The axes are rescaled coordinates $h=(f/\Lambda)^{3/2}H$, $m=(f/\mbox{\ensuremath{\Lambda}})^{3/2}\dot{\pi}$.
The whole phase space is $\left(m,h,\rho\right)$. The deep red regions
are dynamically inaccessible; inside the yellow region the perturbations
of $\pi$ are ghosts. The sandy region has $c_{\text{s}}^{2}<0$ while
in the blue the sound speed is superluminal. The sound speed diverges
as the boundary of the yellow ghosty region. \protect \\
\protect \\
The blue superluminal region to the top right of the yellow ghosty
region is accessible to trajectories and has arbitrarily high sound
speed close to the ghosty region. It appears for arbitrarily small
values of $\rho$ and even at $h=0$. We demonstrate this analytically
in section \ref{sec:superlu}.\protect \\
\protect \\
\protect \\
\protect \\
\protect \\
\protect \\
}
\end{figure}

The situation changes upon adding external matter with positive energy
density $\rho$, coupled to the scalar $\pi$ only through gravity. 
In section \ref{sec:superlu} we have an analytic proof that there are regions in the phase space where the perturbations are propagating faster than light. 
In Fig.~\ref{fig:withrho}, we have plotted a two-dimensional slice
through the now three-dimensional phase space $(m,h,\rho)$, keeping
$\rho$ fixed. Adding external energy density opens up a ghosty region
surrounding the origin of the $(m,h)$ axes, which we have plotted
in yellow. The boundary of this region is a pressure singularity at
which the sound speed diverges. The neighbourhood of this boundary
can emit trajectories and can be approached by trajectories. Therefore
there is always a superluminal region, containing \emph{arbitrarily
high} sound speeds, surrounding this ghosty region. This is true for
arbitrarily small positive external energy densities $\rho$ and for
\emph{all} values of $\beta$ relevant for the \emph{Subluminal Galilean
Genesis} scenario. Adding external energy density allows for new cosmological
configurations of the scalar field $\left(\pi,\dot{\pi}\right)$,
opening up the phase space in this sense. The addition of positive
energy density with a normal equation of state can sometimes transform
a cosmological configuration of the scalar field, into a configuration
without ghosts but with superluminal propagation of perturbations.

Further, we would like to mention that, similarly to the \emph{Ghosts
Condensate }\cite{ArkaniHamed:2003uy} or the simplest \emph{k-inflation}
\cite{ArmendarizPicon:1999rj}, the systems described by (\ref{eq:action})
or (\ref{eq:action_phi}) do not possess any physically acceptable
Lorentz-invariant vacuum. This breakdown is acceptable for a theory
effectively describing a hydrodynamics of an (im)perfect fluid. However,
as a field theory it is far beyond the Standard Model---the QFT which
describes the current state of knowledge in particle physics. The
appearance of ghosts around $X=0$ separates the phase space into
two disconnected sets which correspond to two different EFTs---one
for configurations with spacelike derivatives and another one for
the timelike case. Note that a smooth dynamical evolution from one
set to another, i.e.~from timelike to spacelike derivatives or vice
versa, is physically impossible, because of the pressure singularity
preventing these transitions \cite{Deffayet:2010qz,Pujolas:2011he,Easson:2011zy}
in these theories without a healthy Lorentz-invariant vacuum. This
is obviously the case for the standard representation of fluid dynamics
through fields, see e.g.~\cite{Schutz}. Close to normal vacuum,
where the energy density of the fluid and particle density are zero,
the fluid description should break down in any case and one is forced
to use kinetic theory. In field-theoretical language this breakdown
would correspond to some pathology of the EFT---to an infinitely large
strong-coupling scale or ghosts or unacceptable gradient instability.
This property is crucial for the EFTs which allow for superluminal
propagation. Indeed, if such EFT can only exist for timelike gradients
of some scalar field, then one can always declare this field to be
the time coordinate so that the resulting spacetime is stably causal
and any causal paradoxes are impossible by definition, see e.g.~\cite{Babichev:2007dw}. 

On the other hand, exactly the absence of a physically acceptable
Lorentz-invariant vacuum in the context of an EFT implies that one
cannot use this EFT to describe a transition from inflation (or from
some \emph{Genesis} stage of the early universe as in \cite{Creminelli:2010ba,Creminelli:2012my})
to standard hot Big Bang cosmology. 

Properties of a possible UV completion of an EFT should depend on
\emph{all} states / configurations this EFT is able to describe. It
is not enough to look at a given state or given trajectory and its
small neighbourhood. Indeed, why should the EFT be inapplicable further
away if nothing prevents it from operating there? Only in the case
when the phase space (or Hilbert space) has different unconnected
regions can one claim that there are different EFT's describing these
regions separately. These regions can be separated by states or configurations
with ghosts or gradient instabilities or some classical singularities
or infinitely strong coupling. For example, this is the case for perfect
fluids where the same Lagrangian could be formally used for spacelike
gradients. In some of these configurations with spacelike gradients,
superluminal propagation may be possible. This, however, would not
at all imply that one cannot find weakly coupled particles building
the fluid corresponding to the timelike gradients for which there
is no superluminal propagation. 

In this paper we have only studied external matter which is not coupled
to the \emph{Galileon. }Indeed, this matter should unavoidably be
present in the model. However, we do not think that a non-minimal
direct coupling (\ref{eq:g_matter}) would change the situation. 

Another potential way to avoid our conclusion is to prove that the
energy density $\rho_{\text{s}}\left(\phi,m\right)$ needed to induce
superluminality around a configuration $\left(\phi,m\right)$ brings
the EFT away from its region of validity. However, this is hard to check, since the strong coupling scale $\mu$
(for wave vectors) generically depends on all cosmological phase space
coordinates, so that $\mu=\mu\left(\phi,m,\rho\right)$. On dimensional
grounds, one would expect that the limiting external energy density
for a given $\left(\phi,m\right)$ should be provided by the solution
of the equation $\rho=c_{\text{s}}\left(\phi,m,\rho\right)\mu^{4}\left(\phi,m,\rho\right)$, where $c_\text{s}$ is the sound speed. Unfortunately there is no self-consistent derivation of $\mu$ as
a function of an arbitrary cosmological configuration $\left(\phi,m,\rho\right)$.
This analysis is definitely interesting and important but involves
the calculation of the cubic action for the cosmological perturbations
in this general non-slow-roll setup. This task goes beyond the scope
of our paper. Moreover, it follows from our analysis that the corresponding
$\rho_{\text{s}}\left(\phi,m\right)$ is not necessary parametrically
large. If the addition of such an external $\rho$ can invalidate
the EFT it could imply that reheating might also lie outside
of the region of validity of this EFT.

It would also be interesting to investigate whether one could modify the
theory in such a way that superluminal propagation would be impossible
on \emph{all} configurations with timelike gradients in the presence
of any matter with a normal equation of state. In such a case, the
superluminality could only occur for some unrealistic external matter.
However, given the properties above, this could be impossible and
would definitely be a cumbersome and ambitious task. But only such
strong property could give a hope for a standard Wilsonian UV completion
without a need for something less standard like\emph{ classicalization}
\cite{Dvali:2010jz,Dvali:2010ns}.\\
 \\
We expect that the situation with UV completion is typical for all
recently rediscovered scalar field theories able to violate the NEC. 
\\
\acknowledgments The work of DAE is supported in part by the DOE
under DE-SC0008016 and by the Cosmology Initiative at Arizona State
University. IS is supported by the DFG through TRR33 ``The Dark Universe''.
IS would like to thank the CERN Theory Group for hospitality during
the preparation of this paper. The work of AV is supported by ERC
grant BSMOXFORD no. 228169. AV is thankful to the organisers and staff
of the Cook's Branch Spring Workshop, under the auspices of the Mitchell
Institute for Fundamental Physics and Astronomy at Texas A\&M University
for a warm hospitality during the final stages of this project. 

\bibliographystyle{utphys}
\addcontentsline{toc}{section}{\refname}\bibliography{bounce}

\end{document}